\pdfoutput=1
\documentclass[aps,twocolumn,
floats,prd,nofootinbib,superscriptaddress,10pt]{revtex4-1}
\usepackage[utf8]{inputenc}
\usepackage{bm}
\usepackage{amsmath}
\usepackage{comment}
\usepackage{caption}
\usepackage{color}

\usepackage{graphicx,amsmath,amsfonts,amssymb
}
\def\cyan{\textcolor{black}}

\usepackage{natbib}
\begin{document}

\title{Do we need fine-tuning to create primordial black holes?}

\author{Tomohiro Nakama}
\affiliation{Jockey Club Institute for Advanced Study, 
The Hong Kong University of Science and Technology, Hong Kong, P.R. China}
\author{Yi Wang}
\affiliation{Department of Physics, The Hong Kong University of Science and Technology, 
Hong Kong, P.R.China}
\affiliation{Jockey Club Institute for Advanced Study, 
The Hong Kong University of Science and Technology, Hong Kong, P.R. China}

\begin{abstract}
Recently, the formation of primordial black holes (PBHs) from the collapse of primordial fluctuations has received much attention. The abundance of PBHs formed during radiation domination is sensitive to the tail of the probability distribution of primordial fluctuations. We quantify the level of fine-tuning due to this sensitivity. For example, if the main source of dark matter is PBHs with mass $10^{-12}M_\odot$, then anthropic reasoning suggests that the dark matter to baryon ratio should range between 1 and 300. For this to happen, the \cyan{root-mean-square amplitude of the curvature perturbation} has to be fine-tuned within a $7.1\%$ range. As another example, if the \cyan{recently detected gravitational-wave events are} to be explained by PBHs, the corresponding degree of fine-tuning is $3.8\%$. \cyan{We also find, however, that these} fine-tunings can be relaxed if the primordial fluctuations are highly non-Gaussian, or if the PBHs are formed during an early-matter-dominated phase. We also note that no fine-tuning is needed for the scenario of a reheating of the universe by evaporated PBHs with Planck-mass relics \cyan{left to serve} as dark matter.
\end{abstract}
\maketitle

\section{Introduction}
Primordial black holes (PBHs) with a wide range of masses may have been generated in the early universe (see \cite{Carr:2009jm} and references therein). Among different mechanisms of PBH formation, the collapse of primordial fluctuations during radiation domination is most often discussed in the literature. When estimating the abundance of PBHs created by this mechanism, Gaussianity of primordial fluctuations is often assumed. Under these assumptions the abundance of PBHs is exponentially sensitive to the root-mean-square amplitude $\sigma$ of primordial fluctuations, and hence the generation of a significant amount of PBHs usually involves fine-tuning of $\sigma$, \cyan{posing} a naturalness problem\footnote{See also Ref. \cite{Azhar:2018lzd} for a general discussion of fine-tuning problems with PBH related tuning as an example.}. First we quantify the level of fine-tuning of $\sigma$ required to explain the totality of the cold dark matter by PBHs of $10^{-12}M_\odot$ as an example. We also discuss cases where PBHs of tens of solar masses comprise only part of the entire dark matter, which have recently attracted attention in the context of gravitational-wave astrophysics. 

Two possible ways to \cyan{circumvent} the fine-tuning problem are discussed. First, the large spikes of primordial fluctuations usually accompany with large non-Gaussianities. With the presence of non-Gaussianity, the amount of fine-tuning is reduced. Second, PBHs could have been formed during an early-matter-dominated phase instead of formed during radiation domination. In this case the abundance of PBHs is expressed as power law of $\sigma$ \cite{Khlopov:1980mg,Polnarev:1986bi,Harada:2016mhb}, instead of exponential, which reduces the degree of fine-tuning \cyan{and therefore makes the existence of PBHs more} natural. 

Finally, there is a situation where the universe was reheated by small PBHs, leaving Planck mass relics \cyan{to serve} as dark matter \cite{MacGibbon:1987my,Alexander:2007gj}. In this case, the amplitude of primordial fluctuations has just to be sufficiently large to overproduce PBHs, hence no fine-tuning of $\sigma$ is required.

\section{PBHs formed from Gaussian fluctuations}
\subsection{PBHs formed during radiation domination}
The baryon to photon ratio $\eta$ can be expressed as
\begin{equation}
    \eta\equiv\frac{n_{b,0}}{n_{\gamma,0}}=\frac{\pi^4g_0}{60\zeta(3)}\frac{\rho_{b,0}}{\rho_{r,0}}\frac{kT_{\gamma,0}}{m_pc^2},
\end{equation}
where $\zeta(3)\simeq 1.202$, $g$ represents the effective relativistic degrees contributing the total radiation energy density and the subscripts 0 indicate present values. On the other hand, the fractional energy density $\beta$ of PBHs at some moment during radiation domination can be expressed as
\begin{equation}
\beta=\frac{\rho_{\mathrm{PBH}}}{\rho_r}=\frac{g_0}{g}\left(\frac{g_s}{g_{s,0}}\right)^{4/3}\frac{\rho_{\mathrm{PBH},0}}{\rho_{r,0}}\frac{a}{a_0},
\end{equation}
where entropy conservation has been used and $g_s$ is the effective relativistic degrees contributing to the entropy density. Then the baryon to dark matter ratio can be expressed as
\begin{equation}
    R=\frac{\rho_{b,0}}{\rho_{\mathrm{PBH},0}}=\frac{60\zeta(3)}{\pi^4g_{s,0}}\frac{m_pc^2}{kT}\frac{\eta}{\beta}.
\end{equation}
Suppose $R$ needs to satisfy
\begin{align}
  R_m<R<R_M~.  
\end{align}
Let us consider the case where PBHs of $10^{-12}M_\odot$ comprise all the dark matter \cite{Inomata:2017vxo}, which arose from a sharp spike in the power spectrum of Gaussian primordial fluctuations. The mass of PBHs formed by collapse of overdensities during radiation domination is in proportion to the age of the Universe, and it is $10^5M_\odot$ at $t=1s$ ($T=1$MeV) \cite{Carr:2009jm}. Hence, PBHs of $10^{-12}M_\odot$ formed when $T\sim 10^{8.5}$MeV, neglecting change in $g$ for simplicity, which wouldn't change the discussion here much. 
The quantity $\beta$ for Gaussian fluctuations may be roughly estimated by
$\beta=2^{-1}\mathrm{erfc}(\zeta_c/\sqrt{2}\sigma)$ \cite{Nakama:2017xvq}, with PBH formation threshold chosen as $\zeta_c=0.67$ there. Then, the above condition for $R$ can be rewritten as the condition for $\sigma$ as follows:
\begin{align}
    \!\!\!\!&\!\!\!\!\!\!\!\!\!\frac{\zeta_c}{\sqrt{2}}
    \left[\mathrm{erfc}^{-1}\left(\frac{120\zeta(3)}{\pi^4g_{s,0}}\frac{m_pc^2}{kT}\frac{\eta}{R_M}\right)\right]^{-1}
    <\sigma\nonumber\\
    &\quad\quad\,\,\,\,\,\,
    <\frac{\zeta_c}{\sqrt{2}}
    \left[\mathrm{erfc}^{-1}\left(\frac{120\zeta(3)}{\pi^4g_{s,0}}\frac{m_pc^2}{kT}\frac{\eta}{R_m}\right)\right]^{-1}.
\end{align}
We use $g_{s,0}=3.909$ \cite{Husdal:2016haj}, $\eta=6\times 10^{-10}$ \cite{Steigman:2014pfa}, and also assume $R_m=1/300$ and $R_M=1$ \cite{Tegmark:2005dy}. This value of $R_m$ was estimated from the instability of our galactic disk, so that the disk fragments and star formation takes place. For the value of $R_M$, one may expect $R_M\sim 1$, since in a baryon-dominated Universe matter inhomogeneities on galactic scales are suppressed around the recombination due to Silk damping. This range of $R$ corresponds to $3.4\times 10^{-16}<\beta<1.0\times 10^{-13}$. Then we find $0.0830<\sigma<0.0912$. 

It is convenient to introduce the degree of fine-tuning $\epsilon$ as
\begin{align}
  \epsilon=\frac{\sigma_M-\sigma_m}{(\sigma_M+\sigma_m)/2}. 
\end{align}
When $\epsilon \ll 1$, fine-tuning is needed and the value of $\epsilon$ indicates the amount of tuning needed. When $\epsilon\sim 1$, the scenario can be considered natural. In the example above, $\epsilon\simeq 0.094$, indicating the presence of a fine-tuning. 

\cyan{The above estimation for $\beta$ is simplistic and this issue has recently been revisited} \cite{Yoo:2018kvb}. For a monochromatic spectrum of the primordial curvature perturbation \cyan{they found an improved estimation} $\beta\sim \mu_c^3/\sigma^4\exp(-3\mu_cg_c)\exp(-\mu_c^2/2\sigma^2)$, where $(\mu_c,g_c)=(0.52,-0.141)$. Using this formula we obtain the corresponding range of $\sigma$ as $0.0545<\sigma<0.0585$ ($\epsilon\simeq 0.071$). 

To show how the fine-tuning of $\sigma$ propagates to inflationary model construction, let us consider a specific inflationary scenario\footnote{There is a recent debate that some swampland conjectures may render single-field inflation incompatible with the formation of PBHs \cite{Kawasaki:2018daf}.}, discussed in Ref. \cite{Hertzberg:2017dkh}. The inflaton potential is $U(\phi)=U_0\Sigma_{n=0}^5c_n\phi^n/n!\Lambda^n$, $(c_0,c_2,c_4)=(1,0,1)$. They considered cases where the CMB scale leaves the horizon when $\phi$ is near the origin, and hence the amplitude and the spectral index of large-scale fluctuations are determined by the terms of up to the cubic order. \cyan{Therefore we fix} these terms to match Planck data. As $\phi$ increases, the higher-order terms become important, and depending on $c_5$, perturbations can be enhanced on small scales. They showed there is some value $c_{5,\mathrm{cr}}$, at which small-scale power diverges. This value corresponds to a situation where $\phi$ becomes unable to roll over a small hill in the potential. The deviation $\delta c_5=|c_5-c_{5,\mathrm{cr}}|/|c_{5,\mathrm{cr}}|$ was shown to determine the amount of amplification of small-scale power, or the abundance of PBHs. They found the amplitude ${\cal P}_\zeta(\sim \sigma^2)$ to be ${\cal P}_\zeta/{\cal P}_{\zeta,\mathrm{CMB}}\sim \mathrm{Max}\{10^{-8}C/\delta c_5^2,1\}$, with ${\cal P}_{\zeta,\mathrm{CMB}}\sim 10^{-9}$ and $C\sim 10^{-2}$. Hence, $\sigma$ in this model is determined by $\delta c_{5}$ as $\sigma \sim 10^{-9.5}\delta c_5^{-1}$. 
For simplicity, let us assume Gaussianity and then the range of $\sigma$ we found in the previous paragraph corresponds to $5.40\times 10^{-9}<\delta c_5<5.80\times 10^{-9}$, which is a narrow range and thus the tuning of inflationary parameters is indeed needed. Strictly speaking, one would have to take into account non-Gaussianity in this case as well. We will comment on cases of non-Gaussianity later.

One may also find it useful to see the needed degree of tuning assuming some range for $R$ obtained by recent experiments such as the Planck satellite, instead of using a range for $R$ from an anthropic argument. Let us use $0.182<R<0.192$ \cite{Aghanim:2018eyx}, then we find $0.055563<\sigma<0.055597$, with $\epsilon\simeq 0.00061$. 

PBHs have recently received renewed interests \cite{Bird:2016dcv,Clesse:2016vqa,Sasaki:2016jop} ever since the gravitational-event was detected by LIGO \cite{Abbott:2016blz}. In this case, $f=\Omega_{\mathrm{PBH}}/\Omega_{\mathrm{DM}}$ of much less than unity \cyan{was shown to be} sufficient to account for the gravitational wave events \cite{Sasaki:2016jop}. In this case we \cyan{are unable to} construct an anthropic argument for $\Omega_\mathrm{PBH}$, unlike cases of $f=1$. Nevertheless, observational constraints suggest $10^{-3}<f<10^{-2}$ \cite{Sasaki:2016jop}. For this range of $f$, we use the improved formula for $\beta$ and also a relation $\beta\sim 10^{-8}f$ for PBHs with $\sim 30M_\odot$ \cite{Nakama:2016gzw}. As a result, $0.0624<\sigma<0.0648$ ($\epsilon\simeq 0.038$). If we have used a more conservative range $10^{-4}<f<10^{-1}$, we get $0.0604<\sigma<0.0674$ ($\epsilon\simeq 0.11$). 
\cyan{Note that the sensitivity of $f$ to $\sigma$ is indeed exponential}
: by widening the range of $f$ by two orders of magnitude, the amount of fine-tuning is only relaxed by three times.

\subsection{PBHs formed during an early matter-dominated phase}
PBH formation during an early matter era was originally discussed in Refs. \cite{Khlopov:1980mg,Polnarev:1986bi}, and recently reconsidered in Refs. \cite{Harada:2016mhb,Harada:2017fjm}. Let us again consider PBHs of $10^{-12}M_\odot$, which was formed from fluctuations which reenter the horizon at $t_H\sim 10^{-17}$s. Let $\sigma$ denote the standard deviation of the density perturbation in the linear regime at horizon reentry as in Ref. \cite{Harada:2016mhb}. The scale factor $a_m$ at the average $t_m$ of the moment of maximum expansion satisfies $\sigma a_m/a_H=1$ \cite{Harada:2017fjm}, where $a_H$ is the scale factor at $t_H$. Hence we find $t_m=\sigma^{-3/2}t_H$. PBHs would form at a moment not so later than $t_m$, and note that PBH formation takes place much later than the moment of horizon reentry of fluctuations under consideration, whereas PBHs formed during radiation domination form shortly after the horizon reentry. The abundance of PBHs was found to be $\beta\sim 0.05556\sigma^5$ \cite{Harada:2016mhb}, so the dependence on $\sigma$ is power law instead of exponential, indicating less tuning. This formula was obtained by considering anisotropic collapse, and in addition PBH formation taking into account spins \cite{Harada:2017fjm} and inhomogeneity \cite{kokubu} was also considered. PBH formation from inflaton fragmentation into oscillons was discussed in Ref. \cite{Cotner:2018vug}. See also Refs. \cite{Carr:2017edp,Carr:2018nkm} for PBH formation during an early matter-dominated era. 

Let $t_r$ denote the moment of the reheating, and we need $t_r>t_m$ in order for PBHs to be formed before the end of the early matter-dominated phase. Hence, let us introduce $\gamma(>1)$ and write $t_r=\gamma t_m$. The Friedmann equation at $t_r$ is
\begin{equation}
    H_r^2=\frac{4}{9t_r^2}=\frac{\pi g}{30}T_r^4=\frac{\pi g}{30}t_P^{-2}\left(\frac{T_r}{T_P}\right)^4,
\end{equation}
where $T_r$ is the reheating temperature, $t_P$ is the Planck time and $T_P$ is the Planck temperature. From this one finds
\begin{equation}
    T_r=\left(\frac{40}{3\pi g}\right)^{1/4}\sigma^{3/4}\gamma^{-1/2}\left(\frac{t_P}{t_H}\right)^{1/2}T_P.
\end{equation}
Now we can compute the ratio $R$, introduced above, at the moment of reheating, assuming that the fraction of PBHs in the total matter is constant during the early-matter phase after their formation at $\sim t_m$, to discuss plausible ranges of $\sigma$. 
Whether many more PBHs are formed well after $t_m$ and before $t_r$ or not for a single spike in the primordial spectrum is not well known \cite{Harada:2016mhb}. 

If reheating happens soon after PBH formation, \cyan{we can set} $\gamma= 1$, \cyan{and the range $300^{-1}<R<1$ is translated to} 
$0.00324<\sigma<0.00874$ ($\epsilon \simeq 0.92$), setting $g=106.75$. This shows that the amount of fine-tuning is significantly relaxed. The lower (upper) bound of $\sigma$ here corresponds to $T_r=5.3\times 10^6\, (1.1\times 10^7)$ MeV. The corresponding range for $\beta$ is $2.0\times10^{-14}<\beta<2.8\times10^{-12}$. 

We can also consider larger values of $\gamma$. For instance, when $\gamma=10^{15}$, 
we find $0.0591<\sigma<0.159$ ($\epsilon\simeq 0.92$), with $T_r=2.7\, (5.6)$ MeV, setting $g=10.75$. We observe that the naturnalness is not affected by changing $\gamma$. The corresponding range for $\beta$ is $4.0\times 10^{-8}<\beta<5.7\times 10^{-6}$. 
\section{PBHs formed from non-Gaussian fluctuations}
Often non-Gaussianity of primordial fluctuations is very important in estimating the abundance of PBHs formed by the collapse of non-linear initial fluctuations shortly after horizon reentry during radiation domination \cite{Franciolini:2018vbk}, though Gaussianity is often assumed for simplicity. 
In the following we show that the inclusion of non-Gaussianity can improve the naturalness of PBH production.

First consider a phenomenological model for the non-Gaussian probability density function of the primordial curvature perturbation \cite{Nakama:2016kfq,Nakama:2016gzw,Nakama:2017xvq}:
\begin{equation}
P(\zeta)=\frac{1}{2\sqrt{2}\Tilde{\sigma}\Gamma(1+1/p)}\exp\left[-\left(\frac{|\zeta|}{\sqrt{2}\tilde{\sigma}}\right)^p\right].\label{pdf}
\end{equation}
Here deviations of $p$ from 2 characterize the amount of non-Gaussianity. The mean-square amplitude is 
\begin{equation}
    \sigma^2=\int_{-\infty}^\infty \zeta^2P(\zeta)d\zeta =\frac{2\Gamma(1+3/p)}{3\Gamma(1+1/p)}\tilde{\sigma}^2, 
\end{equation}
where $\Gamma(a)$ denotes the gamma function. The fraction of the universe collapsing to PBHs can be estimated by 
\begin{equation}
    \beta\sim \int_{\zeta_c}^\infty P(\zeta)d\zeta=\frac{\Gamma(1/p,2^{-p/2}(\zeta_c/\tilde{\sigma})^p)}{2p\Gamma(1+1/p)},
\end{equation}
where $\Gamma(a,z)$ is the incomplete gamma function. 
When $p=2$ this formula reduces to the simplistic formula used before, involving the complementary error function, giving 
$0.0830<\sigma<0.0912$ ($\epsilon\simeq 0.094$)
as mentioned before. 

In non-Gaussian cases, when $p=1$, we find
$0.0271<\sigma<0.0324$ ($\epsilon\simeq 0.18$); whereas when $p=0.5$, we find $0.00492<\sigma<0.00684$ ($\epsilon\simeq 0.33$).
That is, the amount of fine-tuning is less severe for smaller values of $p$. This is because $\beta$ is less sensitive to $\sigma$. 

For local non-Gaussianity \cite{Byrnes:2012yx,Nakama:2016gzw,Nakama:2017xvq} $\zeta=\zeta_G+3f_{\mathrm{NL}}/5(\zeta_G^2-\sigma_G^2)$, where $\zeta_G$ is Gaussian, less tuning is needed \cyan{for $f_{\mathrm{NL}}>0$}. We find 
$0.0337<\sigma<0.0374$ ($\epsilon\simeq 0.10$) for $f_{\mathrm{NL}}=10$, $0.0175<\sigma<0.0203$ ($\epsilon\simeq 0.15$) for $f_{\mathrm{NL}}=100$ and $0.0146<\sigma<0.0175$ for $f_{\mathrm{NL}}=1000$ ($\epsilon\simeq 0.18$).
The results become insensitive to $f_{\mathrm{NL}}$ for $f_{\mathrm{NL}}>1000$, since in this case $\zeta$ simply follows a $\chi$-square distribution \cite{Nakama:2016gzw}.

For cubic non-Gaussianity \cite{Nakama:2016gzw,Nakama:2017xvq} $\zeta=\zeta_G+g\zeta_G^3, \quad g\equiv 9g_{\mathrm{NL}}/25$, we find 
$0.0177<\sigma<0.0202$ ($\epsilon\simeq0.13$)
for $g_{\mathrm{NL}}=10^3$, and 
$0.00498<\sigma<0.00660$ ($\epsilon\simeq 0.28$)
for $g_{\mathrm{NL}}=10^{10}$. Again, the result becomes insensitive to $g_{\mathrm{NL}}$ for larger values of $g_{\mathrm{NL}}$ \cite{Nakama:2016gzw}.

\section{Reheating and Planck-mass relics dark matter}
Hawking radiation of small PBHs may leave stable Planck mass relics \cyan{to serve as} cold dark matter \cite{MacGibbon:1987my}. 
If such small PBHs are created during radiation domination and if their mass satisfies $M\lesssim 10^6$g, $\beta$ has to be tiny for relics to account for the dark matter \cite{Carr:2009jm}, which again implies possible fine-tuning. In this case, the energy density of particles emitted as Hawking radiation is negligible. On the other hand, if the initial mass of PBHs is larger, PBHs dominate the Universe before their evaporation, in order for relics to account for the dark matter. In this case, the reheating can also explained by their Hawking radiation \cite{Alexander:2007gj}, for which the initial mass of PBHs has to lie in some interval, as discussed below. 
  \onecolumngrid
\begin{center}
  \begin{table*}[t]
      \begin{tabular}{| c | c | c | c |c|c|c|c|c|}
      \hline
      Case Name & Role      & $M/M_\odot$ & $f=\Omega_{\mathrm{PBH}}/\Omega_{\mathrm{DM}}$                   & $\beta=\rho_{\mathrm{PBH}}/\rho_{\mathrm{r}}$                   &  Era  & Stat. &  $\sigma$ range    & $\epsilon$  \\ \hline
      DMRDG     & All DM    & $10^{-12}$  & 1                   &   $(3.4\times 10^{-16},1.0\times 10^{-13})$                        &  RD   & G     &  $(0.0545,0.0585)$ &0.071       \\ \hline
      GWRDG     & GW events & $30$        & $(10^{-3},10^{-2})$ &    $(10^{-11},10^{-10})$                       &  RD   & G     &  $(0.0624,0.0648)$   & 0.038                \\ \hline
      DMeMDG    & All DM    & $10^{-12}$  & 1                   &     $(2.0\times10^{-14},2.8\times 10^{-12})$                      &  eMD  & G     &  $(0.00324,0.00874)$ & 0.92              \\\hline
      DMRDNG    & All DM    & $10^{-12}$  & 1                   &    $(3.4\times 10^{-16},1.0\times 10^{-13})$                       &  RD   & NG    &  $(0.00492,0.00684)$ & 0.33                 \\\hline
      Quasars       & Quasars   & $10^9$      & $(10^{-11},10^{-9})$         &  $(10^{-15},10^{-13})$    &  RD   & NG    &  $(0.00263,0.00365)$  & 0.32            \\\hline
      \end{tabular}
      \caption{We summarize the naturalness of a few scenarios: (1) PBHs comprising the totality of dark matter, formed in the radiation dominated era with Gaussian initial conditions (DMRDG); (2) PBHs explaining \cyan{the detected binary-black-hole-merger events}, formed in the radiation dominated era with Gaussian initial conditions (DMRDG); (3) Early matter domination with reheating right after PBH formation $\gamma=1$ (DMeMDG); (4) Non-Gaussianities with $p=0.5$ (DMRDNG); and (5) PBHs explaining high-redshift quasars with non-Gaussianities ($p=0.4$).}
      \label{tab:universe}
      \end{table*}
  \end{center}
  \twocolumngrid
 \vspace{-3cm}
  This latter scenario may be more natural, since we just need to overproduce PBHs of some masses. 
  \begin{figure}[t!]
  \includegraphics[width=8.5cm]{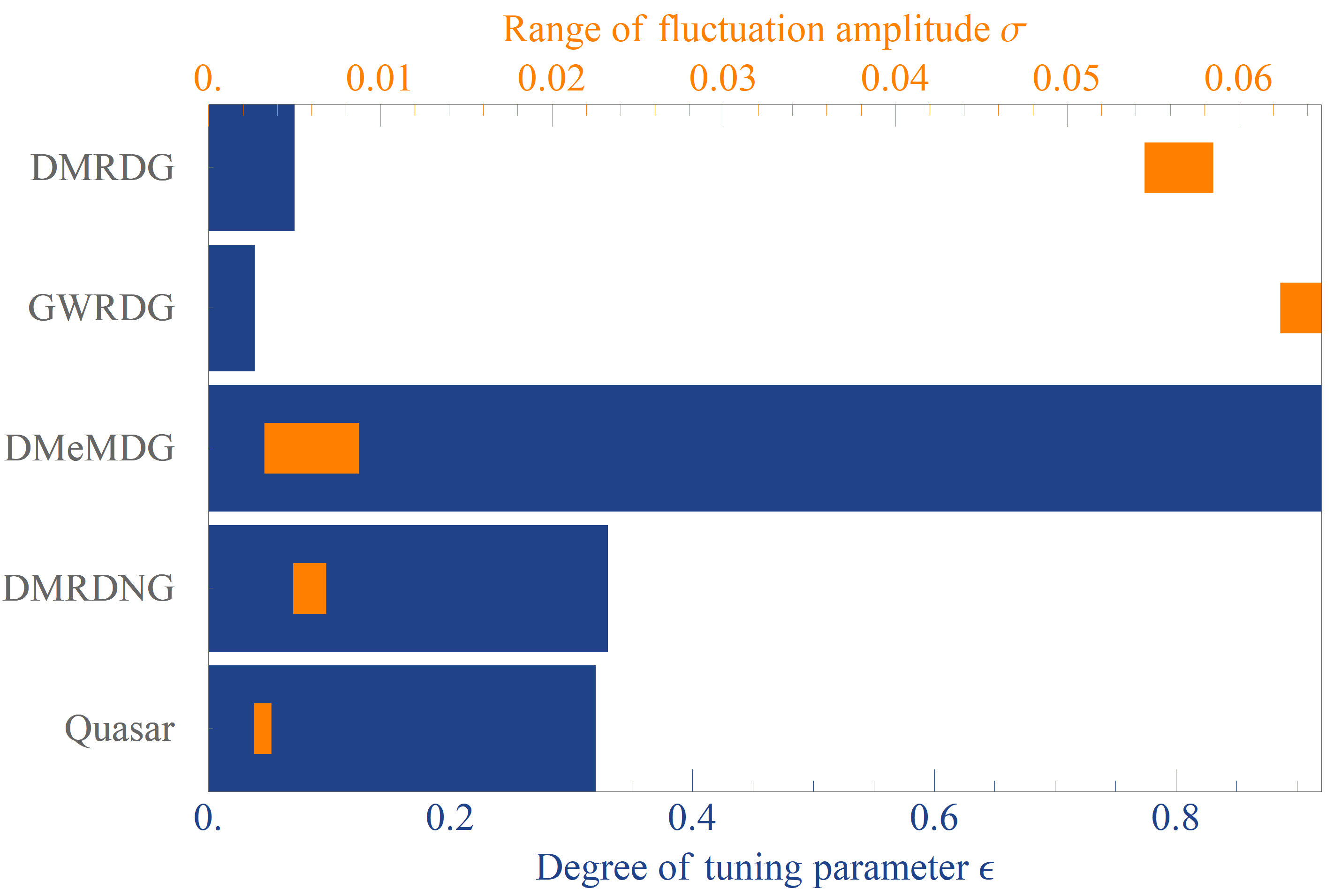}
  \caption{For each case shown in TABLE~\ref{tab:universe}, we plot the degree of fine-tuning $\epsilon$ (lower horizontal axis with blue color) and the required range of $\sigma$ (upper horizontal axis with orange color).}
  \label{fig:universe}
  \end{figure}
  That is, there is no need for fine-tuning of quantities controlling the production efficiency of PBHs, such as $\sigma$ for instance, for PBHs formed from the collapse of primordial fluctuations. The Universe did not have be radiation-dominated at their formation either. 

  Let us assume PBHs with mass $M$ were created abundantly at some point in time, subsequently dominating the early Universe. Their lifetime is \cite{Carr:2009jm} $t_*\simeq 407(M/10^{10}\mathrm{g})^3s\simeq78t_P(M/M_P)^3$, where $t_P$ is the Planck time and $M_P$ is the Planck mass. This was obtained by summing up the contributions of all the standard model particles up to 1 TeV. For simplicity we assume when the age of the universe is $t_*$, these small PBHs evaporated instantaneously to leave Planck mass relics and reheat the universe with temperature $T_*$ and relativistic degrees $g$. The Friedmann equation is 
\begin{align}
    H_*^2&=\frac{4}{9t_*^2}=\frac{8\pi}{3}GMn_*\nonumber\\
    &=\frac{8\pi}{3}t_P^{-2}\frac{M}{M_P}\frac{n_*}{n_P}=\frac{\pi^2g}{30}t_P^{-2}\left(\frac{T_*}{T_P}\right)^4,
\end{align}
where $T_P$ is the Planck temperature and $n_*$ is the number density of PBHs of mass $M$ slightly before $t_*$, which is equal to the number density of relics slightly after $t_*$ under our simplifying assumption. The number density $n_\gamma$ of photons created at $t_*$ is $n_\gamma=2\zeta(3)n_P/\pi^2(T_*/T_P)^3$, so the ratio $R_c=n_*/n_\gamma$ is obtained as $[\pi^3g/160\zeta(3)](M_P/M)(T_*/T_P)$. On the other hand, from the expression for the lifetime $t_*$ in terms of $M$ and the Friedmann equation, one finds $T_*/T_P\simeq0.038(g/106.75)^{-1/4}(M/M_P)^{-3/2}$. Then we find
\begin{equation}
    R_c=A\left(\frac{M}{M_P}\right)^{-5/2},\quad A\simeq0.65\left(\frac{g}{106.75}\right)^{3/4}.
\end{equation}
Suppose $R_c$ needs to satisfy $R_{c,m}<R_c<R_{c,M}$, then $M$ needs to satisfy $(R_{c,M}/A)^{-2/5}<M/M_P<(R_{c,m}/A)^{-2/5}$. Previously we used $R_m<R<R_M$ for the baryon-to-dark-matter ratio, which can be rewritten as $R_M^{-1}\xi_b<\xi_c<R_m^{-1}\xi_b$, where $\xi_b=\rho_b/n_\gamma$ and $\xi_c=\rho_c/n_\gamma$. Here we fix $\xi_b=0.5\times 10^{-28}$ (in Planck unit) \cite{Tegmark:2005dy} for simplicity, and again use $(R_m,R_M)=(300^{-1},1)$. Then we find $1.8\times 10^{10}<M/M_P<1.8\times 10^{11}$ ($\epsilon\simeq 1.6$), \cyan{hence the initial mass of PBHs need not be tuned either.}

\section{Discussion}

We have quantified the amount of fine-tuning in terms of the root-mean-square amplitude of primordial fluctuations. If PBHs comprise the totality of dark matter, an anthropic argument indicates that the primordial fluctuation amplitude has to be fine-tuned. For PBHs comprising a small part of dark matter, no anthropic argument is applicable, while to explain the \cyan{detected binary-black-hole-merger events} using PBHs, a fine-tuning is also needed. In both cases it is assumed that PBHs are formed in the radiation dominated era. Non-Gaussianities or an early matter dominated era can make these scenarios more natural. We summarize these results in TABLE~\ref{tab:universe} and FIG.~\ref{fig:universe}. There we have also included a case where PBHs is responsible for explaining high redshift quasars, \cyan{using Eq. (\ref{pdf}) with $p=0.4$ to avoid an unacceptably large CMB $\mu$ distortion} (see Ref. \cite{Nakama:2016kfq} and references therein). 
Though we have focused on fine-tuning of the amplitude of fluctuations, the scale of enhanced fluctuations, or the moment in time during inflation when enhancement of fluctuations takes place, may also need to be fine-tuned, if one wants to realize a sufficiently narrow mass function of PBHs, to be consistent with different observations.

There \cyan{may be} many other possibilities where natural PBH formation can be achieved without fine-tuning. For example, in \cite{Alexander:2007gj} (see also \cite{Baumann:2007yr,Fujita:2014hha,Hook:2014mla,Hamada:2016jnq}) the authors also proposed the idea of relating small PBHs to baryogenesis. The PBH sector and the baryon sector are thus more closely related and we cannot take $\eta$ and abundance of PBHs as independent parameters. It is interesting to study the naturalness issue in this situation.

\cyan{Furthermore}, we have focused on PBHs formed by collapse of primordial fluctuations, \cyan{but there} are other types of mechanisms of PBH formation, which involve, for instance, vacuum bubbles or topological defects (see Ref. \cite{Carr:2005zd} and references therein). Different mechanisms of PBH formation have different relations between the model parameters involved and the resultant PBH abundance, which implies that the amount of fine-tuning would be different from mechanism to mechanism. One can extend our discussion to such mechanisms, which do not involve collapse of primordial fluctuations, and it would be instructive to evaluate different PBH formation mechanisms from the perspective of the amount of fine-tuning or naturalness.

\begin{acknowledgments}
We thank Jun'ichi Yokoyama for helpful input. The research is supported in part by ECS Grant 26300316 and GRF Grant 16301917 from the Research
Grants Council of Hong Kong. YW would like to thank the participants of the advanced workshop ``Dark Energy and Fundamental Theory" supported by the Special Fund for Theoretical Physics from the Natural Science Foundations of China with Grant No.11747606 for stimulating discussion.
\end{acknowledgments}

\end{document}